# *In situ* $Al_2O_3$ passivation of epitaxial tantalum and aluminum films enables long-term stability in superconducting microwave resonators

Yi-Ting Cheng[1], Hsien-Wen Wan[1], Wei-Jie Yan[2], Lawrence Boyu Young[1], Yen-Hsun Glen Lin[1], Kuan-Hui Lai[1], Wan-Sin Chen[1], Chao-Kai Cheng[1, 3], Ko-Hsuan Mandy Chen[2], Tun-Wen Pi[3], Yen-Hsiang Lin[2,4,*], Jueinai Kwo[2,*], and Minghwei Hong[1,*]

[1]*Grad. Institute of Applied Physics and Dept. of Physics, National Taiwan University, Taipei, Taiwan*
[2]*Department of Physics, National Tsing Hua University, Hsinchu, Taiwan*
[3]*National Synchrotron Radiation Research Center, Hsinchu, Taiwan*
[4]*Taiwan Semiconductor Research Institute, Hsinchu, Taiwan*

## Abstract

Long-term stability of superconducting microwave resonators is essential for scalable quantum technologies; however, surface and interface degradation continue to limit device stability. Here, we demonstrate exceptional stability in microstrip resonators fabricated from epitaxial tantalum and aluminum films, protected by *in situ* deposited $Al_2O_3$ under ultra-high vacuum. These resonators initially exhibit internal quality factors ($Q_i$) exceeding $10^6$ and maintain high performance with minimal degradation after up to fourteen months of air exposure. In contrast, devices relying on native surface oxides show substantial declines in $Q_i$ over time, indicating increased microwave losses. X-ray photoelectron spectroscopy reveals that the *in situ* $Al_2O_3$ effectively suppresses interfacial oxidation and preserves the chemical integrity of the underlying superconducting films, whereas native oxides permit progressive oxidation, leading to device degradation. These findings establish a robust, scalable passivation strategy that addresses a longstanding materials challenge in the development of superconducting quantum circuits.

\* Authors to whom the correspondence should be addressed: yhlin@phys.ntu.edu.tw (Y. H. Lin), raynien@phys.nthu.edu.tw (J. Kwo), and mhong@phys.ntu.edu.tw (M. Hong)

# I. Introduction

Superconducting quantum circuits are among the most promising platforms for quantum computing, where microwave resonators serve as essential components for coupling qubits, enabling quantum state readout, and characterizing circuit performance. These resonators are typically fabricated from heterostructures comprising aluminum (Al) and tantalum (Ta) thin films, whose surfaces rapidly oxidize during air exposure and device processing to form native oxides, such as $AlO_x$ and $Ta_2O_5$[1, 2, 3, 4, 5].

The internal quality factor ($Q_i$) of a resonator, a metric for quantifying microwave loss, depends sensitively on the structural and chemical quality of surfaces and interfaces. Although many superconducting (SC) resonators initially exhibit high $Q_i$ values, prolonged exposure to ambient conditions often results in significant degradation[3, 5]. This deterioration has been linked to chemical inhomogeneities and structural defects introduced by environmental species that diffuse through porous native oxides or migrate along grain and twin boundaries[6]. These defects couple to the microwave fields and act as sources of dielectric loss via mechanisms, such as two-level systems (TLSs)[7, 8].

Considerable efforts have been directed toward improving the initial performance of SC resonators through epitaxial film growth and interface engineering[1, 5, 9, 10]. However, mitigating long-term $Q_i$ degradation remains a key challenge. Conventional wet-etching techniques, such as buffered oxide etch and hydrofluoric acid treatments, can remove native oxides and surface contaminants from Ta resonators[2, 11], but these methods offer limited precision, are incompatible with Al resonators, and cannot prevent reoxidation during subsequent processing steps. Moreover, repeated chemical treatments are impractical for complex, multilayer quantum devices, as they risk damaging sensitive components and are difficult to scale.

Alternative strategies have recently emerged. *In situ* deposition of $Al_2O_3$ on freshly grown ultrathin Al films has effectively preserved pristine film conditions and prevented surface degradation, enabling accurate film characterization[12]. Similar approaches applied to Ta microstrip resonators have shown enhanced stability after air exposure[13]. Another method combines atomic-layer etching (ALE) and atomic-layer deposition (ALD) to replace native $AlO_x$ with ALD-$Al_2O_3$, showing stability in Al-on-Si-based SC quantum circuits[14]; however, this process has not been extended to other superconductors with different surface chemistries. Encapsulation with *in situ* deposition of noble metals such as Au or AuPd also inhibits oxidation[15, 16, 17], but introduces concerns related to proximity effects that may alter superconducting properties.

Despite these advances, a general and scalable solution to preserve SC resonator quality under realistic fabrication and storage conditions remains elusive. To address this challenge, we employ a universal *in situ* passivation strategy in which $Al_2O_3$ films are deposited immediately after epitaxial growth of Al and Ta films under ultra-high vacuum (UHV) conditions. This approach offers three key advantages:

(i) atomically ordered and chemically pristine substrate surfaces;

(ii) abrupt, contamination-free superconductor/substrate interfaces; and

(iii) dense, amorphous $Al_2O_3$ capping layers that protect the heterostructure from environmental degradation during device fabrication and prolonged air exposure.

The *in situ* $Al_2O_3$ passivation acts as a robust diffusion barrier, effectively suppressing oxidation of the underlying SC films. Benchmarking against devices with native oxides, we demonstrate that the deposited-$Al_2O_3$ passivated microstrip resonators based on both Ta and Al films retain $Q_i$ values exceeding $10^6$ after up to fourteen months of air exposure, without post-fabrication treatment. Power-dependent measurements show minimal increases in TLS-related and power-independent losses over time, in stark contrast to devices capped with native oxides. Complementary X-ray photoelectron spectroscopy (XPS) reveals that the *in situ* deposited $Al_2O_3$ films preserve the chemical integrity of the SC films, correlating directly with the high and long-term stability of $Q_i$.

These findings resolve a major bottleneck in superconducting quantum device engineering and establish *in situ* oxide passivation as a practical and scalable strategy for realizing robust, long-lived superconducting circuits for quantum computing technologies

## II. *In situ* deposited $Al_2O_3$ and native oxides on superconducting films - crystallography

To evaluate long-term stability and surface/interfacial chemistry in superconducting microwave resonators, we prepared heterostructure comprising *in situ* deposited $Al_2O_3$ on epitaxial $\alpha$-Ta(110)/sapphire(11$\bar{2}$0) and Al(111)/sapphire(0001), using a custom-built UHV multi-chamber growth and analysis system[18]. For comparison, reference structures of native $Ta_2O_5/\alpha$-Ta(110)/sapphire(11$\bar{2}$0) and native $AlO_x$/Al(111)/sapphire(0001) were also fabricated. Details of substrate preparation, film deposition, and reflection high-energy electron diffraction (RHEED) characterization are provided in the Methods section.

To assess crystallographic quality, we performed high-resolution synchrotron radiation X-ray diffraction (SR-XRD) on Ta and Al films capped with *in situ* deposited $Al_2O_3$ and native oxides. Radial scans along the sapphire$(11\bar{2}0)$ reflection (Figures 1(a), 1(b)) revealed ⟨110⟩-oriented α-Ta films in both the deposited-Al$_2$O$_3$/Ta/sapphire and native Ta$_2$O$_5$/Ta/sapphire. The Ta$(110)$ rocking curves showed narrow full-width-at-half-maximum (FWHM) values of 0.024° and 0.017°, respectively, indicating excellent crystalline quality. The more pronounced Pendellösung fringes in the reference sample suggest a smoother surface and sharper film/substrate interface in that case.

Similarly, radial scans along the sapphire$(0001)$ reflection (Figures 1(c), 1(d)) confirmed ⟨111⟩-oriented Al films in both the deposited-Al$_2$O$_3$/Al/sapphire and native AlO$_x$/Al/sapphire. Narrow rocking curve FWHM values of 0.013° and 0.017°, respectively, demonstrate high crystallinity achieved using both molecular beam epitaxy (MBE) and e-beam evaporation.

In-plane domain symmetries were probed using azimuthal ($\phi$) scans across the Ta$\{101\}$ and Al$\{11\bar{1}\}$ off-normal reflections (Figures 1(e)-1(h)). For Ta films, the $\phi$-scans revealed two sets of fourfold-symmetric diffraction peaks, corresponding to two in-plane domains rotated by 70.5º relative to each other. Each domain exhibited characteristic Ta$\{101\}$ peaks separations of 70.5º, 109.5º, and 70.5º, consistent with the angular relationships expected for the body-centered cubic α-Ta lattice. For Al films, the $\phi$-scans displayed threefold rotational symmetry, with two sets of twin domains rotated by 60º. Each domain displays three diffraction peaks separated by 120º, characteristic of face-centered cubic Al(111).

Notably, the twin domains in Al films had nearly equal volume fractions, independent of growth method. In contrast, the Ta films exhibited unequal domain populations, with one dominant domain and a lower density of domain boundaries. This reduction in interfacial disorder may contribute to the enhanced long-term stability observed in Ta resonators, as discussed in subsequent sections.

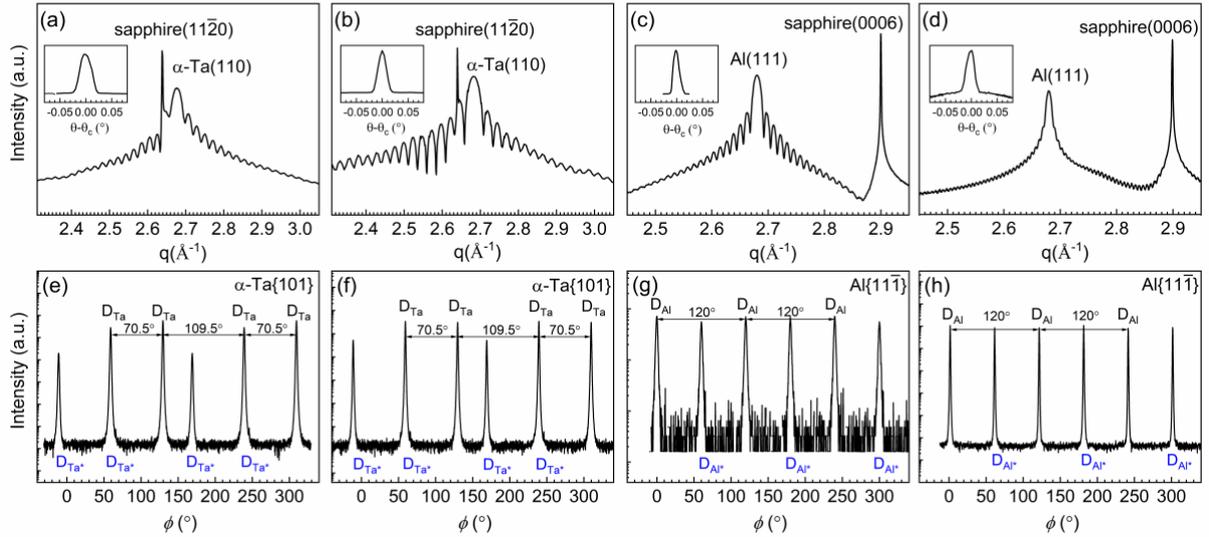

**Figure 1 | High-resolution synchrotron radiation X-ray diffraction (SR-XRD) analysis of epitaxial superconducting films and twin-domain structures.** (a, b) Radial scans along the sapphire$(11\bar{2}0)$ reflection for 30-nm-thick α-Ta films grown on sapphire, capped with (a) *in situ* deposited $Al_2O_3$ and (b) native $Ta_2O_5$. Insets show the corresponding rocking curves of the α-Ta(110) peaks. (c, d) Radial scans along the sapphire(0001) reflection for (c) *in situ* deposited-$Al_2O_3$/50-nm Al/sapphire and (d) native $AlO_x$/100-nm Al/sapphire heterostructures. Insets display the rocking curves of the Al(111) peaks. In (a-d), the abscissa represents the scattering vector, q, calculated as $4\pi \times \sin(2\theta/2)/\lambda$, where $2\theta$ is the scattering angle and $\lambda$ is the X-ray wavelength. (e, f) Azimuthal $\phi$-scans across the α-Ta{101} off-normal reflections, revealing twin-domain structures in α-Ta films. One is denoted as $D_{Ta}$, while the other is referred to as $D_{Ta^*}$. (g, h) Azimuthal $\phi$-scans across the Al{11$\bar{1}$} off-normal reflections, indicating the formation of equally populated twin domains in Al films. One is denoted as $D_{Al}$, while the other is referred to as domain $D_{Al^*}$.

## III. Enhanced long-term stability in superconducting microstrip resonators using *in situ* deposited $Al_2O_3$, in comparison with those using native oxides

### i. Fabrication and measurement setup

Superconducting microwave microstrip resonators were fabricated using heterostructures with either *in situ* deposited $Al_2O_3$ or native surface oxides on epitaxial Ta and Al films. These resonators were designed for coupling to a three-dimensional aluminum superconducting waveguide, enabling precise control of the external quality factor while minimizing packaging-induced losses.

Each resonator was mounted for capacitive coupling to a double-ridge waveguide via a two-end-launcher configuration (Figure 2(a)), which enabled efficient mode conversion from coaxial to waveguide propagation[19, 20, 21]. The waveguide, which does not support a TEM mode, operates over a passband from 3.5 to 10 GHz with an insertion loss of less than 3 dB and provides a flat transmission profile across the relevant frequency range.

Figure 2(b) illustrates the wiring diagram for low-temperature measurements conducted in a dilution refrigerator at a base temperature of 10mK. The transmission coefficient was measured using a vector network analyzer (VNA, ZNB20). Input attenuation of 20 or 60 dB at room temperature, along with an additional 60 dB of cryogenic attenuation, controlled the average photon number <n> in the resonator, ranging from the single-photon regime to approximately $10^7$ photons. The value of <n> is estimated from the room-temperature power at the input channel and the total input-channel attenuation coefficient[5]. The output signal passed through a circulator to a high-electron-mobility transistor (HEMT) amplifier at the 4 K stage, followed by further amplification at room temperature to enhance signal detection.

Figure 2(c) presents a representative transmission coefficient T spectrum of the waveguide from a microstrip resonator fabricated on an *in situ* deposited $Al_2O_3$/Ta/sapphire sample, measured at <n> of 200. Background calibration was performed using spectra acquired above the superconducting transition temperature, where the resonator response is absent. The transmission data were further processed using the fitting procedures described by Probst *et al.*[22], which included corrections for amplitude and phase to accurately extract the $Q_i$.

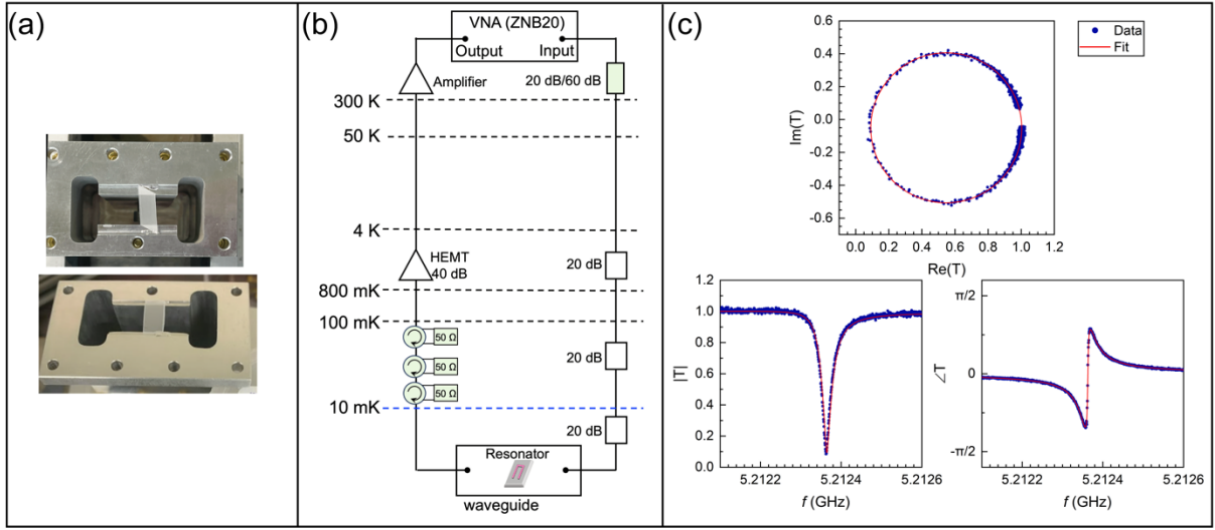

**Figure 2 | Measurement setup and microwave characterization of superconducting microstrip resonators.** (a) Superconducting microstrip resonators based on heterostructures incorporating either *in situ* deposited $Al_2O_3$ or native oxides on Ta or Al films, placed in a 3D waveguide. (b) Schematic of the cryogenic measurement setup used to characterize resonator performance at 10mK in a dilution refrigerator. Input signals are attenuated at multiple stages, and the transmitted signal is amplified via a cryogenic HEMT and room-temperature amplifiers before detection by a VNA. (c) Representative measurement of the transmission coefficient T for a Ta microstrip resonator with *in situ* $Al_2O_3$, shown as: complex-plane response of transmission coefficient T (top) by calibrating the background above $T_c$. The T amplitude as a function of the frequency (bottom left) and phase as a function of the frequency (bottom right). Solid lines indicate fits using a circle-fitting model to extract internal quality factors ($Q_i$).

ii.     **Stability of internal quality factors ($Q_i$) under air exposure**

To assess the long-term stability of superconducting microstrip resonators under air exposure, we measured the $Q_i$ over time for devices fabricated with *in situ* deposited $Al_2O_3$ passivation and native oxides. Measurements were performed immediately after fabrication and again following air exposure periods ranging from several weeks to over a year.

Figure 3(a) presents the $Q_i$ versus average photon number <n> for a Ta microstrip resonator with *in situ* deposited $Al_2O_3$. The as-fabricated device exhibited $Q_i = 1.08 \times 10^6$ at <n> = 6.62. Remarkably, the $Q_i$ remained stable at $0.98 \times 10^6$ at <n> = 6.89 after six months of air exposure, with negligible degradation even after fourteen months. These results confirm that the $Al_2O_3$ passivation layer effectively suppresses the induced losses upon air exposure. The observed loss remains dominated by intrinsic TLS, indicating that no additional dissipation mechanisms were introduced during prolonged air exposure.

In contrast, a Ta resonator capped with native $Ta_2O_5$ exhibited a pronounced decrease in $Q_i$ from $1.35 \times 10^6$ to $0.83 \times 10^6$ after just two months in air, as shown in Figure 3(b). Similar degradation in $Q_i$ has been reported in Ta coplanar waveguide (CPW) resonators with native $Ta_2O_5$[3], highlighting the instability of native $Ta_2O_5$ under air exposure.

A comparable trend was observed in Al resonators, though the degradation was even more severe. As shown in Figure 3(c), Al resonators capped with *in situ* deposited $Al_2O_3$ maintained a stable $Q_i$ above $10^6$ after two weeks of air exposure, with minimal deviation from initial values, indicating minor changes in the TLS-induced loss.

In stark contrast, Al resonators relying on native $AlO_x$ (Figure 3(d)) experienced more than an order-of-magnitude reduction in $Q_i$ over the same period, consistent with previously reported degradation in Al-on-Si-based superconducting resonators[5]. A magnified view in Figure 3(e) reveals that native-$AlO_x$/Al resonators remained power-dependent after two weeks of air exposure.

Taken together, these results demonstrate that *in situ* deposited $Al_2O_3$ passivation significantly enhances the stability of superconducting microstrip resonators. By preventing the formation of defect-rich, unstable native oxides and blocking the diffusion of reactive species, this passivation method preserves low microwave losses over extended timescales. These findings address a longstanding challenge in the development of stable superconducting quantum devices.

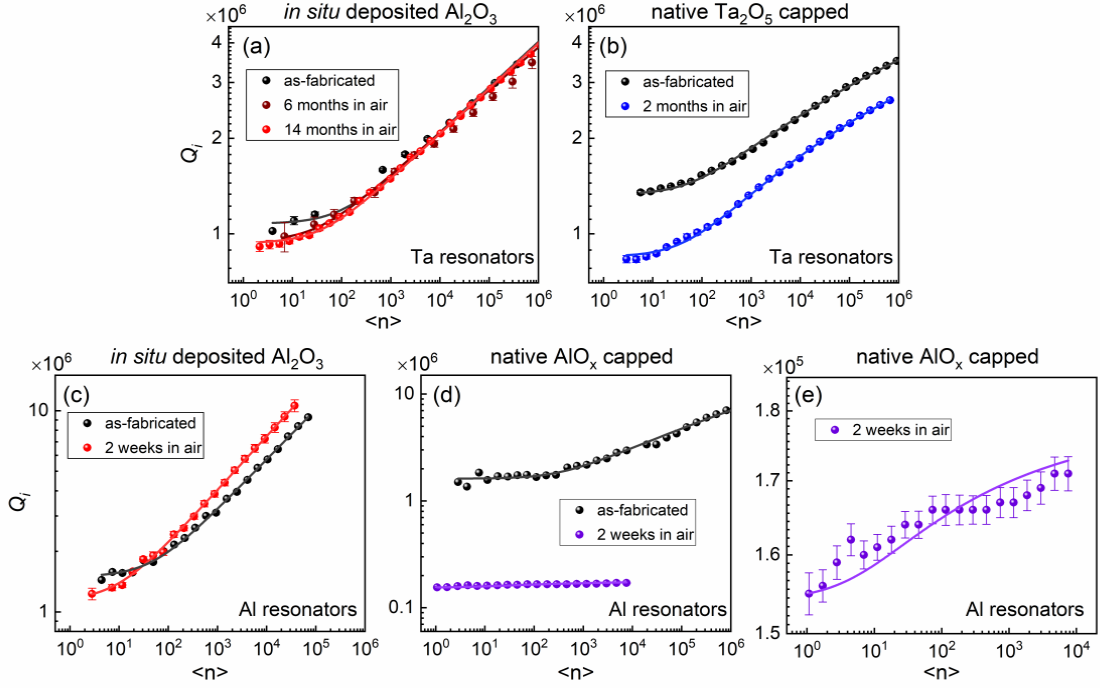

**Figure 3 | Long-term stability of superconducting microstrip resonators with and without *in situ* Al₂O₃ capping.** (a-d) Internal quality factors ($Q_i$) plotted as a function of the applied microwave power, expressed in the average photon number <n> for resonators fabricated from Ta (a, b) and Al (c, d) films. (a) A Ta resonator passivated with *in situ* deposited Al₂O₃ shows negligible degradation in $Q_i$ after prolonged air exposure of six and fourteen months. (b) In contrast, a Ta resonator capped with a native Ta₂O₅ exhibits a decrease in $Q_i$ after just two months in air. (c) An Al resonator passivated with *in situ* deposited Al₂O₃ maintains a stable $Q_i$ after two weeks of air exposure. (d) An Al resonator capped with a native AlO$_x$ shows a one-order-of-magnitude reduction in $Q_i$ over the same period. (e) Magnified view of (d), showing the Al resonator capped with a native AlO$_x$ after two weeks of air exposure. Solid circles represent the measured data, and solid lines are the fitting curves using the model as described in refs.[23, 24].

## IV. Analysis of two-level system (TLS) loss

To elucidate the mechanisms behind the observed long-term stability in $Q_i$, we analyzed the measured resonator data using the following model as described in refs.[23, 24], expressed by:

$$tan\delta = \frac{1}{Q_i} = Ftan\delta_{TLS}^0 \frac{tanh(\frac{hf_r}{2k_BT})}{(1+(\frac{<n>}{n_c}))^\beta} + tan\delta_{other}$$

In this model, $Ftan\delta_{TLS}^0$ denotes the intrinsic low-photon TLS loss, scaled by a geometry-dependent filling factor ($F$); $\beta$ characterizes the TLS saturation behavior due to TLS-TLS interactions[25]; and $tan\delta_{other}$ accounts for power-independent losses such as quasiparticles, vortices, or conductor loss. All analyses were conducted assuming a base temperature of 10mK, consistent with the experimental conditions.

Table I summarizes the extracted loss parameters for resonators with *in situ* deposited $Al_2O_3$ and those with native oxide capping layers, measured immediately after fabrication and after prolonged air exposure.

For resonators protected by *in situ* deposited $Al_2O_3$ films, the intrinsic low-photon TLS-related loss parameter $Ftan\delta_{TLS}^0$ showed only modest increases after long periods of air exposure: a 27.9% increase for Al resonators (from $0.68 \times 10^{-6}$ to $0.87 \times 10^{-6}$) and a 15.1% increase for Ta resonators (from $0.93 \times 10^{-6}$ to $1.07 \times 10^{-6}$). This minimal degradation likely originates from the uncovered sidewalls of the resonators, where the *in situ* deposited $Al_2O_3$ cap does not form a complete barrier.

In stark contrast, resonators capped with native oxides showed substantially larger degradation: For Al resonators, $Ftan\delta_{TLS}^0$ increased by 50.8% (from $0.59 \times 10^{-6}$ to $0.89 \times 10^{-6}$), and Ta resonators experienced a 57.4% increase (from $0.61 \times 10^{-6}$ to $0.96 \times 10^{-6}$). These results demonstrate that *in situ* $Al_2O_3$ passivation effectively suppresses the formation of additional TLS defects induced by air exposure.

In addition to TLS-related loss, we also analyzed the power-independent loss component $tan\delta_{other}$, which remained below $0.03 \times 10^{-6}$ across all *in situ* $Al_2O_3$ passivated devices throughout the testing period, indicating negligible contributions from non-TLS loss channels. However, in Al resonators with native $AlO_x$, $tan\delta_{other}$ increased markedly from $0.03 \times 10^{-6}$ to $5.59 \times 10^{-6}$ after only two weeks of air exposure, indicating the emergence of non-TLS loss channels.

These results underscore the importance of controlled interface engineering in maintaining low-loss superconducting devices. The observed suppression of both TLS and non-TLS loss mechanisms in *in situ* Al$_2$O$_3$ passivated resonators affirms the robustness of this passivation strategy. More broadly, our findings establish a reliable materials pathway for preserving long-term stability in superconducting resonators under realistic environmental conditions, addressing a longstanding challenge in the development of scalable quantum hardware.

**Table I | Summary of resonator properties.** $f_r$: resonance frequency. $Q_c$: coupling quality factor. $Q_{i,LP}$: denotes the internal quality factor at low photon limit. $Ftan\delta_{TLS}^0$: resonator-induced intrinsic TLS loss, which depends on the TLS loss of the system at zero power and temperature ($tan\delta_{TLS}^0$) and geometry through the filling factor ($F$). $tan\delta_{other}$: power-independent loss from various sources. $\beta$: the power dependence factor.

| Microstrip resonator | Time in air | $Q_{i,LP}$ ($\times 10^6$) | $Ftan\delta_{TLS}^0$ ($\times 10^{-6}$) | $f_r$ (GHz) | $Q_c$ ($\times 10^6$) | $tan\delta_{other}$ ($\times 10^{-6}$) | $\beta$ |
|---|---|---|---|---|---|---|---|
| Deposited Al$_2$O$_3$/Ta | 0 | 1.08 | 0.93 | 5.209 | 2.28 | 0 | 0.14 |
| | 6 months | 0.98 | 1.05 | 5.212 | 0.12 | 0 | 0.13 |
| | 14 months | 0.92 | 1.07 | 5.206 | 0.16 | 0 | 0.14 |
| Native Ta$_2$O$_5$/Ta | 0 | 1.35 | 0.61 | 5.075 | 0.70 | 0.14 | 0.15 |
| | 2 months | 0.83 | 0.96 | 5.053 | 0.27 | 0.20 | 0.18 |
| Deposited Al$_2$O$_3$/Al | 0 | 1.44 | 0.68 | 5.126 | 0.42 | 0 | 0.24 |
| | 2 weeks | 1.23 | 0.87 | 5.122 | 0.21 | 0 | 0.26 |
| Native AlO$_x$/Al | 0 | 1.34 | 0.59 | 5.178 | 0.20 | 0.03 | 0.21 |
| | 2 weeks | 0.16 | 0.89 | 5.174 | 0.24 | 5.59 | 0.21 |

### V. Comparative XPS study of Ta and Al films capped with *in situ* deposited Al$_2$O$_3$ and native oxides under air exposure

To elucidate the role of surface passivation in preserving superconducting films upon air exposure, we conducted comparative XPS studies on α-Ta and Al films capped with *in situ* deposited Al$_2$O$_3$, and compared them with corresponding films with native oxides formed via brief air exposure for the Ta film or controlled O$_2$ for the Al film.

For the Ta heterostructure, we examined two types of samples: (i) α-Ta(110) films passivated with *in situ* Al$_2$O$_3$ immediately after growth, and transferred under UHV directly into the XPS chamber; and (ii) a reference sample of α-Ta(110) exposed to air for five minutes to form native $Ta_2O_5$. To evaluate long-term chemical stability, XPS measurements were performed after different durations of air exposure: six weeks and nine months for the deposited-$Al_2O_3$ passivated sample, and one day and one month for the native oxide capped sample.

Figures 4(a) and 4(b) show the Ta 4f core-level spectra, referenced to the metallic Ta peak and normalized in intensity. In the *in situ* deposited-$Al_2O_3$ passivated sample (Figure 4(a)), the line shape of the Ta 4f core-level spectra remains essentially unchanged even after nine months of air exposure, demonstrating that the passivation layer effectively prevents further oxidation of the Ta film. In contrast, the Ta film capped with native $Ta_2O_5$ (Figure 4(b)) shows progressive oxidation with time of air exposure, as evidenced by a growing intensity of $Ta^{5+}$ peaks associated with $Ta_2O_5$[26, 27] and the emergence of a high-binding-energy shoulder attributed to suboxides ($TaO_x$). Prior variable-energy XPS studies[28] have assigned these suboxide features to the interfacial region between the metal and its oxide, which may contribute to dielectric loss in SC resonators.

Similar XPS measurements were carried out on two Al heterostructures: (i) Al(111) passivated with *in situ* deposited $Al_2O_3$, and (ii) Al(111) films exposed to a controlled partial pressure of $O_2$ to form native $AlO_x$. Figures 4(c) and 4(d) show the corresponding Al 2p core-level spectra before and after four days of room air exposure. The *in situ* deposited $Al_2O_3$ sample (Figure 4(c)) shows minimal spectral evolution, affirming the effectiveness of the $Al_2O_3$ passivation in stabilizing surface chemistry. In contrast, the native $AlO_x$-capped sample (Figure 4(d)) exhibits a marked increase in oxidized Al intensity, indicating substantial oxide formation during air exposure; this may stem from patchy and incomplete oxidation of the initial controlled $O_2$ exposure, which leaves regions of bare metallic Al to further oxidation upon exposure to air. Although a slight increase in intensity of the oxidized Al peak is observed in the $Al_2O_3$-passivated sample, likely due to incomplete coverage over the Al film, the increase is comparatively minor.

These XPS results correlate well with the resonator performance data. Al resonators with *in situ* deposited $Al_2O_3$ maintain stable $Q_i$ over time, whereas those with native $AlO_x$ caps degrade rapidly, underscoring the critical role of surface chemistry in long-term device stability.

This issue is particularly critical for superconducting metals such as Ta and Al films grown on sapphire substrates, where twin boundaries and surface defects act as preferential oxidation pathways. Native oxides tend to be porous and defective, allowing environmental species to diffuse through the film. Along the domain boundaries, these reactive species may eventually reach the superconductor-substrate interface, which typically features a high participation ratio of electric field. This diffusion exacerbates dielectric loss and accelerates performance degradation. In contrast, the dense and chemically inert nature of the *in situ* deposited $Al_2O_3$ blocks such diffusion pathways, preserving the chemical integrity of the underlying superconducting films.

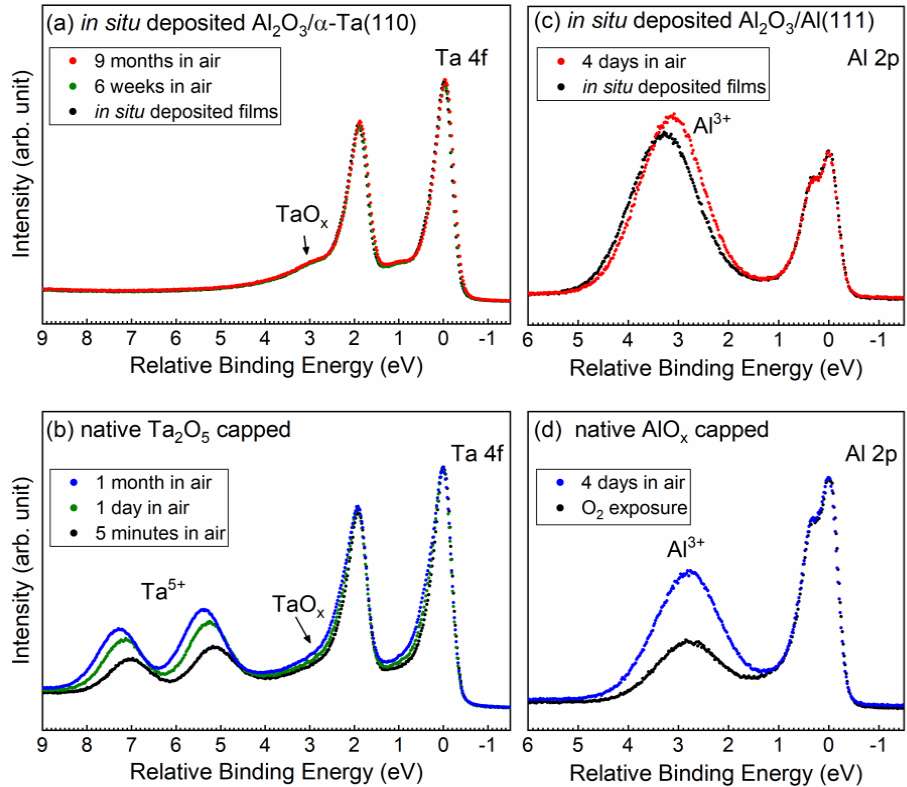

**Figure 4 | X-ray photoelectron spectroscopy (XPS) analysis of Ta and Al films with and without *in situ* $Al_2O_3$ passivation upon air exposure.** (a) Ta 4f core-level spectra of an *in situ* deposited $Al_2O_3/\alpha$-Ta(110) films, showing minimal spectral changes between the as-deposited surface and after prolonged air exposure of six weeks and nine months, indicating effective protection by the deposited $Al_2O_3$ thin film. (b) Ta 4f core-level spectra of a $\alpha$-Ta(110) film upon air exposure of five minutes, one day, and one month, revealing progressive oxidation of the Ta surface. (c) Al 2p core-level spectra of an *in situ* deposited $Al_2O_3/Al(111)$ films before and after four days of air exposure, demonstrating stable surface passivation with negligible evolution in chemical composition and minimal native aluminum oxides formation. (d) Al 2p core-level spectra of an Al(111) film with native $AlO_x$ capped, before and after four days of air exposure, showing significant thickening of the surface aluminum oxides after air exposure. In (c, d), the oxidized Al peak is associated with $Al^{3+}$.

## VI. Conclusion

Realizing practical quantum computing requires long-term stability of superconducting circuits, making materials passivation a critical challenge. We have demonstrated exceptional resistance to aging in superconducting resonators fabricated with *in situ* deposited $Al_2O_3$ films directly grown on pristine Ta and Al surfaces under ultra-high vacuum. These devices initially exhibited $Q_i$ values exceeding one million, and retained this performance with minimal degradation even after extended air exposure lasting up to fourteen months. Rapid degradation of $Q_i$ values was observed in the resonators capped with native oxides. Our systematic XPS analyses revealed that the *in situ* deposited $Al_2O_3$ effectively passivates the underlying superconducting films, preventing them from further reactions with environmental species. In contrast, upon exposure to air, progressive oxidation was observed in Ta and Al films capped with their respective native oxides.

The strategy of engineering *in situ* deposited oxide on freshly grown superconducting films stabilizes the superconducting films against environmental degradation and provides a solution for a longstanding materials bottleneck in the development of robust and scalable superconducting quantum hardware.

## VII. Methods

**Preparation of atomically ordered and chemically clean sapphire substrates**

Analyzing dielectric losses at various interfaces concludes that imperfections at the SC film/substrate interface significantly increase TLS losses, which have been mitigated through improved substrate cleanness[2, 10, 12, 29, 30, 31, 32, 33]. Sapphire substrates with a-plane ($11\bar{2}0$) or c-plane ($0001$) orientations were chemically cleaned using a piranha solution, rinsed with deionized water, and dried with high-purity nitrogen gas. The substrates were then loaded into the UHV multi-chamber system and thermally annealed at 800°C under a base pressure of $3 \times 10^{-10}$ Torr. The resulting surfaces exhibited sharp, streaky, and bright reflection high-energy electron diffraction (RHEED) patterns with Kikuchi arcs, as shown in Figures S1(a) and S1(b) for the processed sapphire($11\bar{2}0$) and sapphire($0001$) substrates, respectively, indicative of atomically smooth, well-ordered, and contamination-free surfaces, as confirmed by *in situ* X-ray photoelectron spectroscopy (XPS).

**Growth of Al$_2$O$_3$/α-Ta(110)/sapphire(11$\bar{2}$0)**

Epitaxial α-phase Ta(110) films, 30-nm-thick, were immediately deposited on the above-described a-plane sapphire(11$\bar{2}$0) substrates via DC magnetron sputtering at 700°C. The sputtering was performed at a base pressure of 1.7 × 10$^{-9}$ Torr and an argon pressure of 3.75 × 10$^{-3}$ Torr, with applied powers of 175 W and 50 W. Note that the sputtering chamber is UHV connected to the multi-chamber system. The *in situ* RHEED patterns acquired after Ta deposition showed bright and sharp streaks along the ⟨1$\bar{1}$1⟩ azimuth (Figure S1(c)), indicating epitaxial growth of an α-Ta(110) film with high crystallinity. The deposited Ta films exhibited a superconducting transition at 4.3K, consistent with bulk α-Ta[34]. After Ta film deposition, the samples were transferred *in situ* to the oxide chamber, cooled to near room temperature (to ensure that the deposited Al$_2$O$_3$ films were amorphous), and deposited with nominally 2 nm thick Al$_2$O$_3$. The non-crystalline nature of the deposited Al$_2$O$_3$ on Ta and Al film was revealed with *in situ* RHEED study, shown in Figures S1(e) and S1(f), respectively.

**Growth of Al$_2$O$_3$/Al(111)/sapphire(0001)**

The procedure for Al film growth via molecular beam epitaxy (MBE) was previously described[12]. An effusion cell provided a growth rate of 0.045 nm/s, and the substrate temperature was nominally below 0°C. The RHEED image of a 50 nm-thick Al film (Figure S1(d)) revealed bright and sharp streaks along the ⟨2$\bar{1}\bar{1}$⟩ and ⟨0$\bar{1}$1⟩ azimuths of Al(111), confirming epitaxial growth on sapphire(0001). After deposition, the samples were *in situ* transferred to the oxide chamber and capped with nominally 3 nm thick amorphous Al$_2$O$_3$.

**Preparation of native Ta$_2$O$_5$/α-Ta(110)/sapphire(11$\bar{2}$0)**

Following deposition of the 30-nm-thick α-Ta(110) films, wafers were cooled for three hours, then exposed to ambient air. Oxide thicknesses of 1.2 nm, 2.15 nm, and 2.64 nm were obtained after air exposure for five minutes, one day, and one month, respectively. Thicknesses were estimated from the Ta 4f oxide-to-metal peak ratio using the method in ref.[35]. RHEED images of the native oxide surfaces, Ta$_2$O$_5$ and AlO$_x$, shown in Figures S1(g) and S1(h), respectively, exhibited diffuse, halo-like features consistent with amorphous surface layers.

**Preparation of native AlO$_x$/Al(111)/sapphire(0001)**

For fabricating an Al resonator consisting of native AlO$_x$/Al/sapphire(0001), the growth started with wafer cleaning, followed by loading the cleaned wafer into a *Plassys* electron beam evaporator. A 100-nm-thick Al film was deposited at a rate of 5.78 Å/s at a base pressure in the mid-$10^{-7}$ Torr range. The Al surface was then oxidized under a static oxygen pressure of 7.6 Torr for ten minutes at room temperature. The resulting native AlO$_x$ layer thickness was estimated to be 2.79 nm based on the oxide/metal XPS peak intensities. For XPS studies, Al films were grown via MBE[12] in a UHV multi-chamber growth/analysis system[18]. The films were *in situ* transferred to the load-lock for oxidation under the same static O$_2$ pressure, duration, and temperature conditions, then transferred to the XPS chamber.

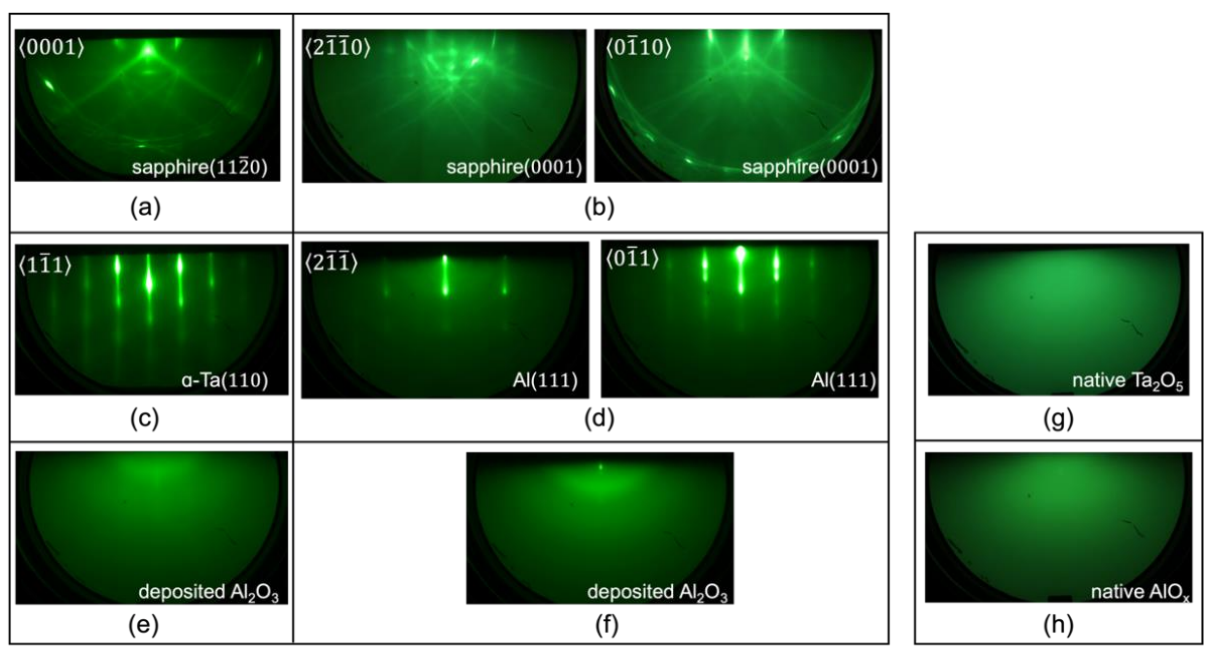

**Figure S1 Reflection high-energy electron diffraction (RHEED) patterns** of (a) sapphire (11$\bar{2}$0) along the ⟨0001⟩ azimuth, (b) sapphire(0001) along the ⟨2$\bar{1}\bar{1}$0⟩ and ⟨0$\bar{1}$10⟩ azimuths, (c) a 30-nm-thick α-Ta(110) film along the ⟨1$\bar{1}$1⟩ azimuth, (d) a 50-nm-thick Al(111) film recorded along the ⟨2$\bar{1}\bar{1}$⟩ and ⟨0$\bar{1}$1⟩ azimuths, (e, f) *in situ* deposited amorphous Al$_2$O$_3$ on Ta and Al films, respectively, (g) a native Ta$_2$O$_5$ film formed by air exposure, and (h) a native AlO$_x$ film formed by O$_2$ oxidation.

**Synchrotron radiation X-ray diffraction**

Synchrotron X-ray diffraction (SR-XRD) was employed to evaluate film crystallinity and structure. Measurements were performed at BL17B and BL09A beamlines of Taiwan Light Source (TLS) and Taiwan Photon Source (TPS), respectively, at the National Synchrotron Radiation Research Center (NSRRC), Taiwan, using 9.5 keV incident X-ray at room temperature.

**Resonator fabrication**

Resonators were patterned using photolithography and etched via inductively coupled plasma-reactive ion etching (ICP-RIE). AZ5214E photoresist was spun at 5800 rpm for 100 seconds for an approximate resist thickness of 1.3 - 1.4 μm and soft baked at 100°C for 2 minutes. Patterning was performed using a mask aligner equipped with near-ultraviolet (NUV) light exposure and TMAH-based developer, followed by rinsing in deionized (DI) water.

Etching employed a $CF_4$:$O_2$ (10:1) gas mixture for Ta stacks (deposited-$Al_2O_3$/Ta and native $Ta_2O_5$/Ta) and a $BCl_3$:$Cl_2$ (1:2) mixture for Al stacks (deposited-$Al_2O_3$/Al and native $AlO_x$/Al). Ta resonators were exposed to $O_2$ plasma post-etching in the same system to form a designed $Ta_2O_5$ layer on the sidewalls. Al resonators underwent $CF_4$ etching to remove $AlCl_x$, followed by DI water rinsing to dilute any residual $AlCl_x$ after being taken out from the etching system.

Photoresist was removed in 1-methyl-2-pyrrolidone (NMP) at 80°C for 30 min (Ta resonator) or 60 min (Al resonator) with ultrasonic agitation. Samples were diced using a diamond scribing wheel into 0.5 cm × 1.5 cm parallelograms (Ta resonator) and 0.5 cm × 1.1 cm rectangles (Al resonator). Resonators with a linewidth of 10 μm were designed following Zoepfl et al.[36], targeting the resonance frequency in the range of 5.0-5.2 GHz.

**Acknowledgment**

We thank Prof. Alp Sipahigil (University of California, Berkeley) and Prof. David I. Schuster (Standford University) for valuable discussions. The authors thank the support from the National Science and Technology Council (NSTC) through No. NSTC 113-2119-M-007-008-. Y. H. Lin is supported by the Ministry of Education YuShan Young Scholar Fellowship.

**Author contributions**

M. Hong, J. Kwo, and Y.H. Lin supervised the project. Y.T. Cheng and H.W. Wan carried out the epitaxial growth of Ta films and the e-beam evaporation of $Al_2O_3$ thin films on Ta. Y.H.G. Lin and W.S. Chen contributed to the growth of epitaxial Al films and the deposition of $Al_2O_3$ films on Al. C.K. Cheng and Y.T. Cheng conducted synchrotron radiation X-ray diffraction experiments. Y.T. Cheng fabricated the Ta resonators. K.H. Lai and L.B. Young contributed to the fabrication of Al resonators. Y.H. Lin and W.J. Yan designed the resonator geometry and 3D waveguide, and set up the low-temperature measurement system. W.J. Yan performed the low-temperature microwave measurements. W.J. Yan and Y.T. Cheng analyzed the microwave measurement data. Y.T. Cheng carried out X-ray photoelectron spectroscopy (XPS) measurements on the Ta samples and performed data analysis for all XPS experiments, while W.S. Chen conducted the XPS measurements on the Al samples. K.H.M. Chen performed transport measurements on the Ta samples. T.W. Pi provided insights and comments on the XPS data. All authors contributed to reviewing and editing the manuscript.